\documentclass{article}
\usepackage[T1]{fontenc}

\makeatletter

\newcommand{\LyX}{L\kern-.1667em\lower.25em\hbox{Y}\kern-.125emX\@}

\usepackage{epsfig,wrapfig,amssymb}
\makeatother

\begin{document}

\title{Studying Short-Range Dynamics in Few-Body Systems}

\author{J. A. Templon\thanks{
This work supported by the U.S. National Science Foundation under grant NSF-PHY-9733791.
}\\
\emph{Department of Physics and Astronomy, The University of Georgia}\\
\emph{Athens, Georgia USA 30602}}

\date{12 February 1999}

\maketitle
\thispagestyle{empty}

\begin{abstract}
I present the case for studying the nature of short-range internucleon interactions
with electron-scattering experiments on few-body nuclear targets. I first review
what electron-scattering studies have unearthed about the nature of the interactions
between nucleons in nuclei at small separation. Special consideration is given
to a couple of recent experiments. The results essentially serve to construct
a roadmap for future studies in this area. The related experimental program
at Jefferson Lab is presented, along with suggestions for future theoretical
work. 
\end{abstract}
The most important length scale characterizing nuclei is roughly \( \hbar /m_{\pi }c \)
\cite{fria98}. This is no accident, since exchange of pions is responsible
for the most important part of the interaction between nucleons. There is another
important length scale, but it is normally less apparent than the first. This
scale is the nucleon radius, and it is important since nucleons are observed
to repel each other strongly when their separation becomes less than 1 fm. It
is not normally so apparent since the classical nuclear-physics literature is
usually expressed in terms of the independent-particle model, which itself can
be derived via Hartree-Fock type calculations using effective interactions.
These interactions, however, have to be generated by a procedure which takes
the strong short-range interaction into account. Thus in a real sense, those
interactions are essentially responsible for the observed properties of nuclei.

The internucleon interaction at a length scale of \( \hbar /m_{\pi }c \) is
well understood in terms of the exchange of physical mesons. At shorter range,
this interaction is phenomenological. This presumably reflects a breakdown of
the meson-exchange picture at small separations, and our inability to carry
out QCD calculations at low energies. It is interesting to measure this important
component of the internucleon interaction, as we expect it will tell us something
about how nuclear interactions evolve from mesonic to chromodynamic degrees
of freedom.

\section{The Case for Few-Body Systems}

Intermediate-energy electron scattering is the tool most suited to mapping the
properties of individual nucleons in a nuclear medium \cite{day90,lapi93,kell96,pand97}.
For heavier systems, theoretical calculations must use techniques which build
the anticorrelation between nucleon locations, due to the short-range repulsion,
into the strength of the interaction. The advantage of using a few-body system
for the target is that the \( NN \) interaction is directly used for the computation
of the wave functions. For \( A=3 \) systems, Faddeev techniques allow a direct
computation of the spectral function \cite{kiev97}, and for heavier light nuclei,
the technique of integral transforms \cite{leid98} can be used to construct
the spectral function. The spectral function \( S(E_{m},p_{m}) \) is closely
related to the electron-scattering cross section and provides a probability
distribution of nuclear protons versus their momentum \( p_{m} \) and binding
energy \( E_{m} \).

\section{Results from Inclusive Electron Scattering}

\begin{wrapfigure}[15]{l}{72mm}  \epsfig{file=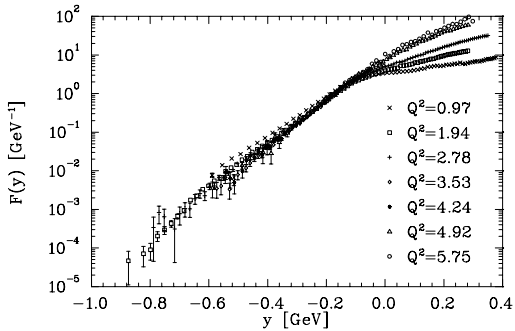,width=67mm}
  \caption{Scaling function derived from Jefferson Lab \((e,e')\) data.}
  \label{fig:arrington}
\end{wrapfigure} At intermediate energies and quasifree kinematics, many inclusive
\( (e,e') \) experiments have been performed. Plane-wave reasoning suggests
that at large \( Q^{2}=-q^{2} \), the \( (e,e') \) cross section should become
a function of only two factors. The first is the incoherent cross section to
scatter electrons from all the nucleons in the nucleus, and the second is a
partial integral \( F(y) \) over the proton spectral function \cite{atti91}.
\( y \) is essentially the component of the struck proton's momentum along
the \( (e,e') \) momentum transfer \( \vec{q} \); it is also closely related
to the deviation of \( \omega =E_{e}-E_{e'} \) from the quasielastic value
\( \omega \approx |\vec{q}|^{2}/2m_{N} \).

Figure \ref{fig:arrington} shows the most recent \( (e,e') \) data from Jefferson
Lab \cite{arri98}. \( F(y) \) is constructed as 
\begin{equation}
\label{eq:fy}
F(y)={d^{2}\sigma \over d\Omega d\omega }[Z\sigma _{ep}+N\sigma _{en}]^{-1}{q\over (M^{2}+(y+q)^{2})^{\frac{1}{2}}}
\end{equation}
 For \( y<0 \) (low \( \omega  \) relative to the quasielastic peak), data
for different kinematics are in excellent agreement, indicating that the effects
beyond PWIA are small.

In studying short-range phenomena, access to specific regions in \( S(E_{m},p_{m}) \)
is desirable so data on \( F(y) \) are not sufficient. Coincidence data are
required to access these regions. However there is one further inclusive measurement
of interest, namely that of the Coulomb Sum Rule. This sum rule relates the
energy-integrated longitudinal response from \( (e,e') \) to the proton-proton
correlation function \cite{defo83b}. However, analyses have so far been inconclusive
due to large theoretical corrections for reaction effects (\emph{e.g.} meson-exchange
currents (MEC)) and for incomplete \( \omega  \) coverage in the experiments
\cite{carl98}.

\section{Coincidence \protect\( (e,e'p)\protect \) experiments}

Coincidence \( (e,e'p) \) experiments can in principle more directly probe
the spectral function \( S(E_{m},p_{m}) \). In plane wave, the cross section
is 
\begin{equation}
\label{eq:cpwia}
\frac{d^{6}\sigma }{d\Omega _{e'}dE_{e'}d\Omega _{p}dE_{p}}=|\vec{p}_{p}|E_{p}\sigma _{ep}S(E_{m},p_{m})
\end{equation}
 and an extraction of the spectral function is unambiguous. The variables \begin{wrapfigure}[21]{l}{72mm}   \epsfig{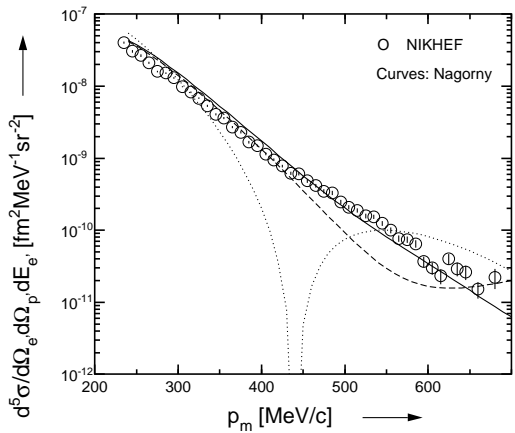}
  \caption{\(^{4}\mbox{He}(e,e'p)^{3}\mbox{H}\) cross section measured at
    NIKHEF.  The dotted curve is a PWIA calculation, and the other
    two curves include various classes of additional reaction effects.}
  \label{fig:nagorny}
\end{wrapfigure} \( E_{m} \) and \( p_{m} \) are computed by using the measured
four-momenta of the incident electrons, scattered electrons and knocked-out
protons to reconstruct the four-momentum of the residual \( (A-1) \) system
\( R=(E_{R},\vec{p}_{R}) \). \( p_{m}=|\vec{p}_{R}| \) and \( E_{m}=\sqrt{R^{2}}+m_{p}-M_{A} \).
However, additional reaction-mechanism effects can break the direct link between
the cross section and spectral function.

Fig.~\ref{fig:nagorny} shows data measured at NIKHEF \cite{leeu98b} for the
reaction \( ^{4}\mbox {He}(e,e'p)^{3}\mbox {H} \). The dotted curve is the
plane-wave prediction, and the sharp minimum is a feature of the spectral function
which has been directly linked to the short-range part of the \( NN \) interaction
\cite{tado87}. The data do not exhibit this minimum, and the calculation attributes
this discrepancy to \( p \)--\( t \) final-state interactions (FSI) and to
MEC.

Another example of reaction effects thwarting access to interesting information
comes from the large-\( E_{m} \) data from the same experiment. Simple arguments
lead to the prediction \cite{kest96,ciof96} of a ``ridge'' in the spectral
function, due to short-range \( NN \) interactions, along the locus \( E_{m}\sim 2S_{N}+(p_{m})^{2}/2m_{N} \)
where \( S_{N} \) is the single-nucleon separation energy. Computations of
the spectral function have supported this prediction.

\begin{wrapfigure}[28]{l}{42mm}  \epsfig{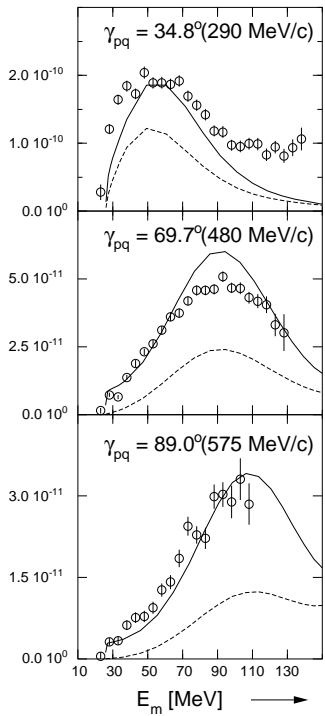}
  \caption{Large-\(E_{m}\) data for \(^{4}\mbox{He}(e,e'p)\) from NIKHEF.
    The mean \(p_{m}\) for each pane is indicated.}
  \label{fig:kvl-cont}
\end{wrapfigure}

Fig.~\ref{fig:kvl-cont} shows data for \( ^{4}\mbox {He}(e,e'p) \) at large
\( E_{m} \) \cite{leeu98a} along with theoretical predictions. The peak in
the cross section (for both the data and the curves) follows the ridge relation
noted above. However, the theory indicates that only about half of the observed
cross section is due to direct knockout (dashed line). The rest is due to MEC.
Also, for the lowest-\( p_{m} \) data (the top pane), the calculation severely
underpredicts the data at large \( E_{m} \).

\section{The Few-Body Program at Jefferson Lab}

 The preceding discussion makes clear that accessing the spectral
function in regions of \( (E_{m},p_{m}) \) relevant to short-range nuclear
dynamics is difficult. The problem is that the spectral function is relatively
much smaller in these regions than at lower momenta and energies. This leads
to the possibility that other reaction processes, even if weak, can substantially
contaminate the data.

Many ideas have been formulated about how to suppress these contaminant processes
in experiments. These ideas were difficult to implement in experiments at labs
such as NIKHEF and Mainz, mainly because their beam energies were too low to
provide the necessary kinematic flexibility. I now discuss some of these ideas
and how they are being implemented at Jefferson Lab.

\subsection{Parallel Kinematics}

\begin{wrapfigure}[10]{r}{84mm}   \epsfig{file=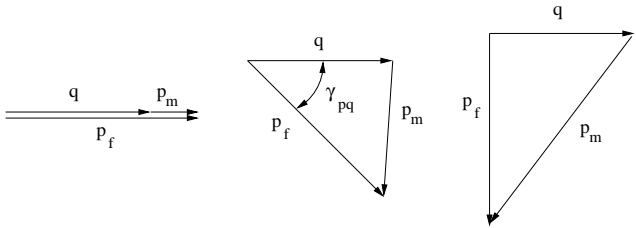,width=84mm}
  \caption{Various values of \(p_{m}\) for fixed \((\omega, \vec{q})\).}
  \label{fig:parperp}
\end{wrapfigure}

Fig.~\ref{fig:parperp} depicts how measurements (including those of \cite{leeu98b,leeu98a})
of cross sections at large \( p_{m} \) were previously made. The simultaneous
constraints on \( \omega  \), \( q \), and \( p_{m} \) made it impossible
to reach large \( p_{m} \) values unless the knocked-out protons were detected
at large angles with respect to \( \vec{q} \). Elastic FSI can seriously distort
measurements in this type of measurement, since at the same electron kinematics,
reactions such as that at left in Fig.~\ref{fig:parperp} are also possible.
The associated spectral function is several orders of magnitude larger due to
the lower \( p_{m} \) involved. Such low-\( p_{m} \) protons can rescatter
through large angles and contribute to (and perhaps even dominate) the large-\( p_{m} \)
cross section. This qualitative argument is supported by calculations \cite{fran97}
which show that FSI contributions are at a minimum in parallel kinematics. 

The large beam energies available at Jefferson Lab make it possible to perform
large-\( p_{m} \) experiments at parallel kinematics, and several proposals
utilizing this principle \cite{sick97,zhao97,temp97} are already on the books.

\subsection{Variation of \protect\( Q^{2}\protect \)}

Experiments at lower-energy labs were not able to make substantial variations
in \( Q^{2} \) for a given \( (E_{m},p_{m}) \) region. \( Q^{2} \) variations
are useful in two respects: to help discriminate between one- and two-body currents
contributing to the cross section; and to suppress the contaminant (two-body)
currents. The one-body direct-knockout process of interest only depends on \( Q^{2} \)
through the electron-proton cross section, while MEC and IC contributions are
expected to have a very different \( Q^{2} \) behaviour. There is disagreement
about whether larger or smaller \( Q^{2} \) experiments are better for suppressing
the two-body currents; \( (e,e') \) analyses appear to favor smaller \( Q^{2} \),
but the difference is only significant for \( y\gtrsim 0 \) (see Fig.~\ref{fig:arrington}).
All of the experiments studying short-range dynamics at Jefferson Lab plan to
make measurements at multiple values of \( Q^{2} \).

\subsection{Large Negative \protect\( y\protect \) Values}

In Fig.~\ref{fig:arrington} the data clearly violate the scaling hypothesis
for \( y>0 \). This is generally accepted to result from contributions outside
the one-body impulse approximation framework. At negative values of \( y \),
the data scale well. In addition, theoretical studies \cite{fran97} have indicated
that FSI are best suppressed when the ejected proton's longitudinal (along \( \vec{q} \))
component is large and negative; this condition also yields a large, negative
\( y \) value. I should mention that these studies indicate that FSI are also
suppressed when the longitudinal momentum is large and positive, but it is unclear
how this condition constrains two-body currents. Two experiments in Hall A at
Jefferson Lab \cite{zhao97,temp97} plan to make measurements at large negative
\( y \) kinematics.

\subsection{Suppression of Multistep FSI}

Ingo Sick has pointed out \cite{sick97} an additional mechanism which contaminates
\( (e,e'p) \) measurements at large \( E_{m} \). Multistep FSI, or \( p \)--\( N \)
scattering within the nucleus, change both the energy and direction of knocked-out
protons. This causes the proton to be detected with \( (E_{m},p_{m}) \) values
much different than those at the \( (e,e'p) \) reaction vertex. If these FSI
``move'' events from a region where the spectral function is large to a region
where it is low, these ``moved'' events can generate cross sections larger
than the ``native'' protons at this \( (E_{m},p_{m}) \) which did not undergo
multistep FSI.

\begin{wrapfigure}[18]{r}{7cm}    \epsfig{file=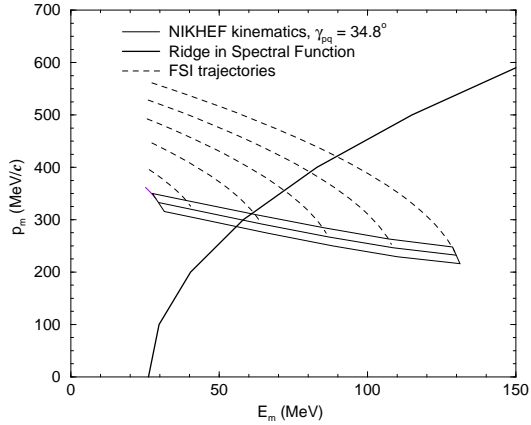,width=7cm}
    \caption{FSI trajectories for NIKHEF large-\(E_{m}\) data.}
  \label{fig:kvl-fsi-traj}
\end{wrapfigure} Fig.~\ref{fig:kvl-fsi-traj} shows the kinematics in the \( (E_{m},p_{m}) \)
plane for the upper pane of Fig.~\ref{fig:kvl-cont}. The dark line shows the
``ridge'' in the spectral function where the greatest strength is expected.
The dashed lines show how multistep FSI move events in the \( (E_{m},p_{m}) \)
plane; reactions with vertex \( (E_{m},p_{m}) \) values all along the dashed
lines can contribute, by undergoing a \( (p,p'N) \) reaction, to the experimental
measurement (the box is the experimental acceptance, and the thin solid line
gives the central kinematics for which these calculations were performed.) It
is clear that for missing energies greater than about 65 MeV, one may expect
increasing contributions from multistep FSI to the data. This is a plausible
explanation for the calculation's underprediction of the data for \( E_{m}>90 \)
MeV.

An approved experiment \cite{sick97} in Hall C will make measurements on both
sides of the ``ridge'' and in several different types of kinematics, to test
whether this effect is indeed important.

\section{Representative Expected Results at Jefferson Lab}

Fig.~\ref{fig:97-111} shows an example of what we hope to achieve at Jefferson
Lab. This figure is a calculation for \( ^{4}\mbox {He}(e,e'p)^{3}\mbox {H} \)
at a beam energy of 4 GeV. Experiment \cite{temp97} in Hall A proposes to measure
this reaction in an attempt to observe the spectral-function minimum discussed
in relation to Fig.~\ref{fig:nagorny}. The dashed lines in Fig.~\ref{fig:97-111}
are plane-wave calculations; the solid curves include FSI in the framework of
the Generalized Eikonal Approximation \cite{fran97}. The upper curve corresponds
to \( y\approx 100 \) MeV/\emph{c}, and the bottom curve corresponds to \( y\approx 400 \)
MeV/\emph{c}. The bottom curve is also computed for parallel kinematics. The
calculations display the expected reduction in FSI due to parallel kinematics
and large \( -y \) values. Unfortunately, these calculations do not yet include
two-body current contributions.

\section{Outlook}

\begin{wrapfigure}[22]{r}{7cm}    \epsfig{file=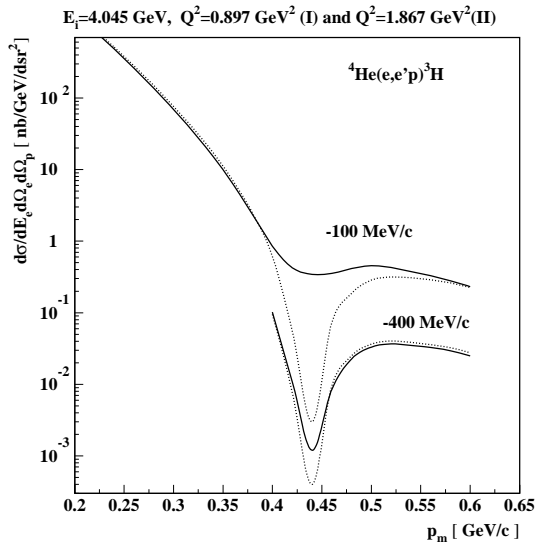,width=7cm}
    \caption{Expected results for experiment 97-111 in Hall A.}
  \label{fig:97-111}
\end{wrapfigure}

A broad program exists to study \( (e,e'p) \) reactions, with an emphasis on
few-body nuclei, at Jefferson Lab. The experiments comprising this program have
a new set of tools, courtesy of the large JLab beam energy, with which to (attempt
to) force nature to give us clean information about the nuclear spectral function
in regions relevant to short-range nuclear dynamics. Parallel kinematics will
be an important feature of almost all these experiments. Furthermore, data will
be taken at a variety of \( Q^{2} \) and \( y \) settings in an attempt to
suppress two-body current contributions to a manageable level.

I have not mentioned two other powerful techniques which will be exploited at
Jefferson Lab and elsewhere: response-function separations and multi-nucleon
knockout experiments. Both techniques are in principle more selective for accessing
the large-momentum one-body current of interest. However, both are experimentally
more demanding, thus the program outlined above provides a better starting point
for testing our understanding of the \( (e,e'p) \) reaction mechanism at high
energies. The results can be used to design more effective response-function
separation or multinucleon-knockout experiments.

On the theoretical side, there are many nice frameworks, models and techniques
in circulation for computing spectral functions exactly, treating two-body currents,
computing FSI at large proton momenta, and so on. However, no one group seems
to have all ``nice'' ingredients. Figure \ref{fig:97-111} provides a good
example; it uses a state-of-the-art spectral function from the Argonne group,
and a modern FSI computation, but no two-body currents are included. It is highly
unlikely that the program outlined above will suppress reaction effects to the
point that PWIA is valid; interpretation of these results will require close
collaboration with our theoretical colleagues.

\section*{Acknowledgements}

The author wishes to thank Dr. E.~Jans (NIKHEF), spokesperson for the NIKHEF
experiment discussed here, for a critical reading of this manuscript and useful
suggestions.

\bibliographystyle{prsty}
\bibliography{mainz}

\end{document}